\documentclass[reprint,amsmath,amssymb,aps,pra]{revtex4-1}
\usepackage{graphicx}
\usepackage{bm}
\usepackage{braket}
\usepackage{mathtools}
\usepackage{amsfonts}
\usepackage{amsmath}
\usepackage[breaklinks,colorlinks=true,linkcolor=red,urlcolor=blue,citecolor=magenta]{hyperref}

\begin{document}
\title{Flat bands and higher-order topology in polymerized triptycene: \\
Tight-binding analysis on decorated star lattices}
\author{Tomonari Mizoguchi}
\affiliation{Department of Physics, University of Tsukuba, Tsukuba, Ibaraki 305-8571, Japan}
\email{mizoguchi@rhodia.ph.tsukuba.ac.jp}
\author{Mina Maruyama}
\affiliation{Department of Physics, University of Tsukuba, Tsukuba, Ibaraki 305-8571, Japan}
\author{Susumu Okada}
\affiliation{Department of Physics, University of Tsukuba, Tsukuba, Ibaraki 305-8571, Japan}
\author{Yasuhiro Hatsugai}
\affiliation{Department of Physics, University of Tsukuba, Tsukuba, Ibaraki 305-8571, Japan}

\newcommand{\txtr}[1]{\textcolor{red}{#1}}
\newcommand{\txtrout}[1]{\textcolor{red}{\sout{#1}}}
\newcommand{\txtb}[1]{\textcolor{blue}{#1}}
\newcommand{\txtg}[1]{\textcolor{green}{#1}}
\newcommand{\txtm}[1]{\textcolor{magenta}{#1}}

\newcommand{\mvecthree}[3]
{
  \left[
    \begin{array}{c}
#1  \\
#2  \\
#3  
    \end{array}
    \right]
}
\date{\today}
\begin{abstract}
In a class of carbon-based materials called polymerized triptycene,
which consist of triptycene molecules and phenyls, 
exotic electronic structures such as Dirac cones and flat bands 
arise from the kagome-type network. 
In this paper, we theoretically investigate the tight-binding models for polymerized triptycene,
focusing on the origin of flat bands and the topological properties. 
The mechanism of the existence of the flat bands is elucidated by using the ``molecular-orbital"
representation, which we have developed in the prior works.
Further, we propose that the present material is a promising candidate to 
realize the two-dimensional second-order topological insulator, which is characterized by the 
boundary states localized at the corners of the sample. 
To be concrete, we propose two methods
to realize the second-order topological insulator, and 
elucidate the topological properties of the corresponding models
by calculating the corner states as well as  
the bulk topological invariant, namely 
the $\mathbb{Z}_3$ Berry phase.
\end{abstract}

\maketitle
\section{Introduction}
Carbon-based materials have been a fertile ground for realization of exotic phenomena in condensed matter physics. 
Graphene~\cite{Novoselov2004}, a monolayer of carbon atoms forming a honeycomb lattice, is a typical example,
where exotic transport phenomena and magnetic responses arise from the massless Dirac dispersion~\cite{Wallace1947,CastroNeto2009}. 
The characteristic edge modes~\cite{Fujita1996,Okada2001,Peres2006} 
and topological properties~\cite{Ryu2002,Novoselov2006,CastroNeto2006,Hatsugai2006} have also been of great interests. 

In graphene, carbon atoms are networked through $sp^2$-orbitals, leading to the 
nearest-neighbor (NN) tight-binding description on a honeycomb lattice for the remaining $\pi$ orbital.  
Recently, it has been found that richer geometric and electronic structures can be achieved when considering 
hybrid networks of $sp^2$ and $sp^3$ carbon atoms~\cite{Maruyama2013,Maruyama2016,Sorimachi2017,Maruyama2017,Fujii2018,Fujii2018_2,Shuku2018}.
In the present paper, we focus on the class of materials called polymerized triptycene. 
The basic constituents of these materials are
triptycene molecules~\cite{Bartlett1942} (Fig.~\ref{Fig:trip} A), 
a member of iptycene family~\cite{Chen2013}, 
and phenyl groups interconnecting the neighboring triptycene molecules (Fig.~\ref{Fig:trip} B).
Synthesis of this family of materials has actively been pursued~\cite{Chong2007,Zhang2012,Zhang2013,Liang2013,Bhola2013,Zhang2015,Liang2018}.
In the theoretical side, the first principles calculations 
have revealed the characteristic band structure of this class of materials~\cite{Fujii2018_2}, 
namely, $\pi$ electrons on $sp^2$ hydrocarbons form kagome-type network, 
which supports the multiple flat bands with surprisingly good flatness,
as well as the massless Dirac dispersion around K and K$^\prime$ points.

The goal of this paper is to present a deeper understanding of the electronic structures
and to pursue the possibility of topological phases in a series of polymerized triptycene, by means of the tight-binding analysis
on decorated star lattices. 
The main focus of the previous studies~\cite{Maruyama2013,Maruyama2016,Sorimachi2017,Maruyama2017,Fujii2018,Fujii2018_2} is 
the emergence of flat bands, which opens up a way to intriguing many-body effects such as the flat-band ferromagnetism~\cite{Mielke1991,Tasaki1992,Tasaki1998}. 
In this context, we elucidate the origin of flat bands in the present model by using the ``molecular-orbital" (MO) representation,
which is a generic framework to describe the flat-band models~\cite{Hatsugai2011,Hatsugai2015,Mizoguchi2019}.

In addition to the flat-band physics, 
we also propose that the present material is a promising candidate for the novel topological phase of matter
proposed recently, namely, the higher-order topological insulator (HOTI). 
The HOTI has the characteristic boundary states, 
protected by the topological nature of Bloch wave functions in the bulk, 
at the boundaries with co-dimension larger than one~\cite{Hayashi2018,Benalcazar2017,Song2017,Schindler2018, Ezawa2018,Xu2017,Schindler2018_2,Khalaf2018,Kunst2018,Ezawa2018_2,Ezawa2018_3,Hayashi2019,Kang2018,Araki2019,Calugaru2019,Agarwala2019}.
Recently, it was found that the kagome-lattice model is a promising platform to realize the HOTI.
The key ingredient is the ``breathing" of the lattice structure, i.e., 
the modulation of the size of triangles  (or the amplitude of hoppings on those triangles) 
in such a way that the upward triangles are larger or smaller than the downward ones~\cite{Ezawa2018,Kunst2018,Xu2017,Araki2019}. 
Indeed, the HOTI in the breathing kagome lattice is experimentally realized 
in various metamaterials, such as photonic crystals~\cite{Ohta2019,Hassan2019}, phononic crystals~\cite{Xue2019,Ni2019}
and a (111) surface of Cu with molecular scatterers~\cite{Kempkes2019}.
Motivated by these theoretical and experimental studies, 
we propose that polymerized triptycene is a suitable platform to realize the kagome-based HOTI in the solid-state system.
To demonstrate this, we analyze the corner states in the finite sample of 
polymerized triptycene under the open boundary condition, and relate them to a bulk topological invariant for HOTIs, namely, 
quantized Berry phase~\cite{Hatsugai2011,Berry1984, Hatsugai2006_2,Kariyado2018,Kawarabayashi2019,Kudo2019,Araki2019_2}. 

The rest of this paper is organized as follows. 
In Sec.~\ref{sec:kagomenet}, 
we present the characteristic geometric structure of polymerized triptycene 
and explain how the kagome-type network is formed. 
In Sec.~\ref{sec:model}, 
we present tight-binding models which describe a series of polymerized triptycene,
and elucidate their characteristic band structures, such as flat bands and Dirac cones. 
Understanding of band structures from the viewpoint of the kagome bands
is also explained. 
In Sec.~\ref{sec:hoti}, 
we discuss the possible realization of the HOTI phase in this material.
In Sec.~\ref{sec:discussion}, 
we discuss the perspectives of this class of materials beyond the tight-binding analysis, 
such as the stability of the HOTI phase against disorders and the effects of correlations. 
In Sec.~\ref{sec:summary}, we present a summary of this paper. 
In Appendix~\ref{sec:mo}, 
we explain how to determine the flat-band energies and 
present the explicit form of the MO representation of the tight-binding model.

\section{Kagome-type network of polymerized triptycene \label{sec:kagomenet}}
\begin{figure}[b]
\begin{center}
\includegraphics[width= 0.95\linewidth]{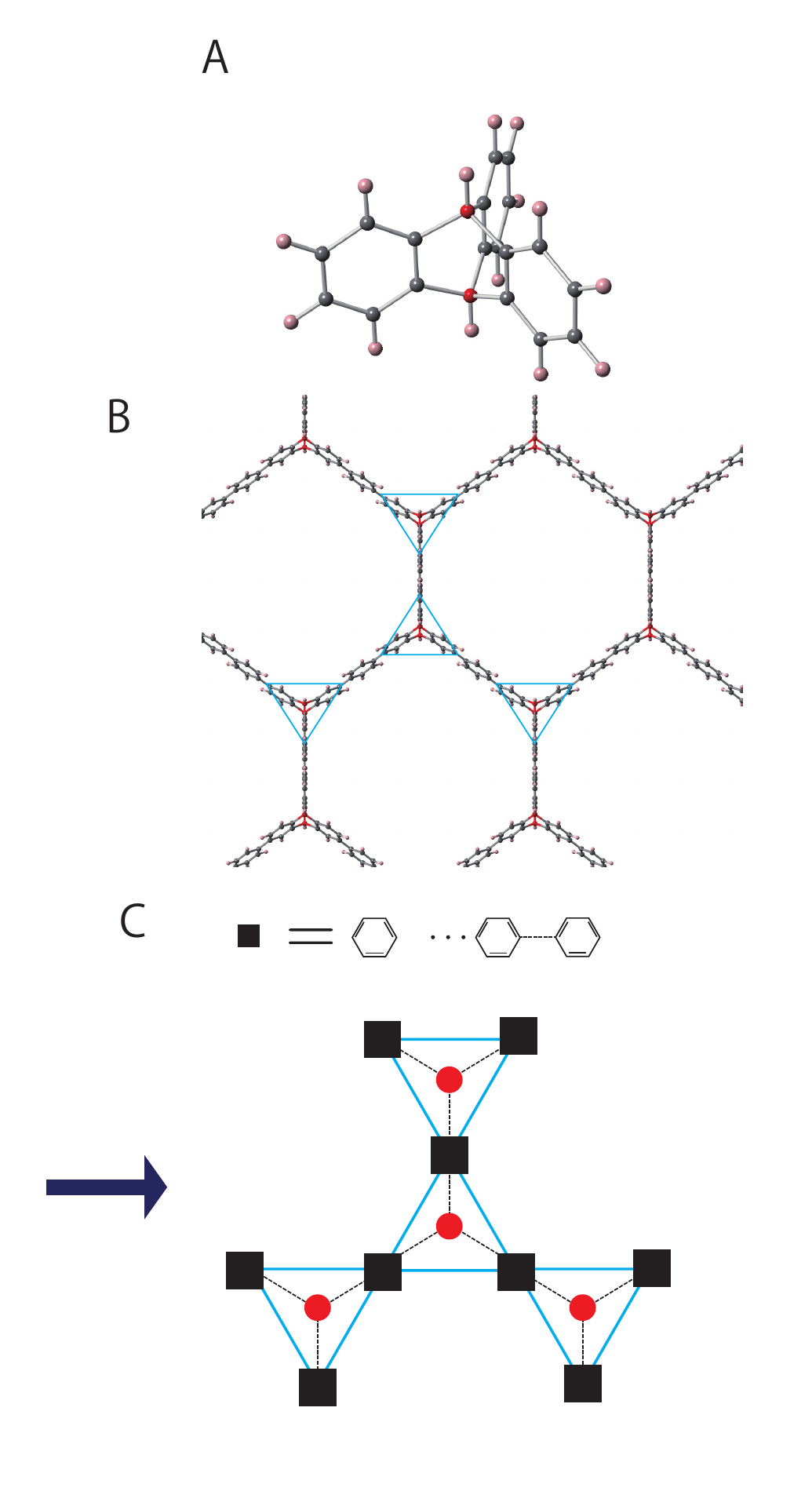}
\caption{A: The structure of a tripticene molecule.
Red spheres 
denote the $sp^3$ carbon atoms.
B: Polyermized triptycene containing one phenyl between neighboring triptycene molecules.
Blue triangles denote triptycene molecules.
C: Schematic figure for the polymerized triptycene and its kagome network obtained by reducing chains of C$_6$ rings as a single site.}
\label{Fig:trip}
\end{center}
\end{figure}
A triptycene molecule is a family of iptycene molecule; here the prefix ``tri" means that it contains three C$_6$ rings (Fig.~\ref{Fig:trip} A). 
Remarkable feature of this molecule is that there exist two kinds of carbon atoms:
One has $sp^3$ electric configuration which interconnects three C$_6$ rings (red spheres or dots in Fig.~\ref{Fig:trip}),  
and the other has $sp^2$ electric configuration which composes C$_6$ rings.

Since three C$_6$ rings on each triptycene molecule form 120-degree structure, 
it is possible to construct a two-dimensional network of triptycene molecule by placing them on a honeycomb lattice and connecting the neighboring C$_6$ rings 
by other molecules, such as acenes~\cite{Fujii2018} or phenyls~\cite{Fujii2018_2}.  
We call the materials thus obtained as polymerized triptycene. 
In the present paper, we consider the materials containing phenyls. 

The schematic figure for the polymerized triptycene is shown in Fig.~\ref{Fig:trip} B.
Here the carbon atoms with $sp^3$ configuration are denoted by red dots. 
Since the triptycene molecules are placed on a honeycomb lattice, 
the chains of C$_6$ rings are placed on its dual lattice, namely, a kagome lattice.
In other words, if each chain of C$_6$ rings is regarded as a single ``molecule",
the molecular orbitals of them are aligned on a kagome lattice (Fig.~\ref{Fig:trip} C).
This is a geometrical origin of the kagome-type network in this class of materials. 

In the following, we analyze the tight-binding models
where each C$_6$ ring is replaced by a single site (Figs.~\ref{Fig1} C \ref{Fig4} A, and \ref{Fig4} C). 
Such tight-binding models provide a simplified description of electronic structures
in a sense that six carbon atoms are reduced to one site.
Nevertheless, they are useful to capture the characteristic electronic structures 
arising from the kagome network, as demonstrated in the previous work based on the first-principles calculations~\cite{Fujii2018_2}.

\section{Tight-binding Hamiltonian \label{sec:model}}
\subsection{Polymerized triptycene containing phenyl}
\begin{figure}[b]
\begin{center}
\includegraphics[width= 0.95\linewidth]{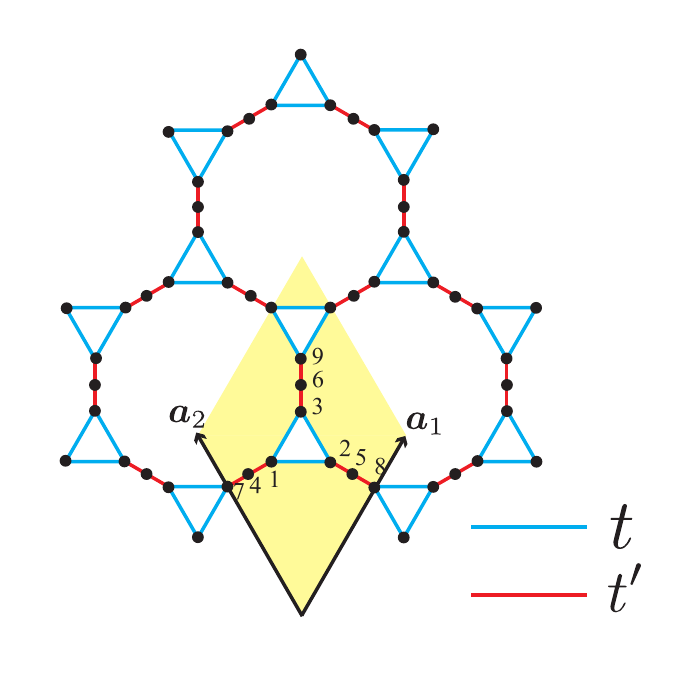}
\caption{The tight-binding model of Eq. (\ref{eq:ham}).
The primitive vectors and the sublattices are indicated in the figure. }
\label{Fig1}
\end{center}
\end{figure}
In what follows, for simplicity, we neglect the spin degrees of freedom of the electrons.
The role of spin degrees of freedom will be discussed in Sec.~\ref{sec:discussion} in the context of correlation effects. 

For the polymerized triptycene containing one 
phenyl between neighboring triptycene molecules, the tight-binding Hamiltonian 
(see Fig.~\ref{Fig1} for its schematic figure) is written as 
\begin{equation}
H^{\mathrm{P}}= \bm{c}_{\bm{k}}^\dagger \mathcal{H}^{\mathrm{P}}_{\bm{k}} \bm{c}_{\bm{k}}, \label{eq:ham_tp}
\end{equation}
where $\mathcal{H}^{\mathrm{P}}_{\bm{k}}$ is a $9\times 9$ matrix,
\begin{equation}
\mathcal{H}^{\mathrm{P}}_{\bm{k}} = \left[
\begin{array}{ccc}
t h^{\bigtriangleup} & t^\prime E_3 & 0  \\
t^\prime E_3  & 0  & t^\prime E_3   \\
0 &  t^\prime E_3  & t h^{\bigtriangledown}_{\bm{k}}  \\
\end{array}
\right], \label{eq:ham}
\end{equation}
with 
\begin{eqnarray}
h^{\bigtriangleup} = \left[
\begin{array}{ccc}
0  & 1 & 1 \\
1 & 0 & 1 \\
1 & 1 & 0 \\
\end{array}
\right],
\end{eqnarray}
and
\begin{eqnarray}
h^{\bigtriangledown}_{\bm{k}} = \left[
\begin{array}{ccc}
0  & e^{-i \bm{k}\cdot (\bm{a}_1-\bm{a}_2)} & e^{-i \bm{k}\cdot \bm{a}_1}  \\
e^{i \bm{k}\cdot (\bm{a}_1-\bm{a}_2) } & 0 & e^{-i \bm{k}\cdot \bm{a}_2} \\
e^{i \bm{k}\cdot \bm{a}_1} & e^{i \bm{k}\cdot \bm{a}_2}  & 0 \\
\end{array}
\right].
\end{eqnarray}
Note that $E_n$ represents a $n\times n$ identity matrix,
and
$\bm{c}_{\bm{k}} = \left(c_{\bm{k},1} , c_{\bm{k},2} , c_{\bm{k},3} , c_{\bm{k},4} ,c_{\bm{k},5} ,c_{\bm{k},6},
c_{\bm{k},7} ,c_{\bm{k},8} ,c_{\bm{k},9} 
\right)^{\rm T} $
represents the annihilation operators of electrons with momentum $\bm{k}$, written in a form of a nine-component vector. 

The lattice structure is similar to that of a star lattice~\cite{Zheng2007,Yang2010,Lee2018}, 
where the vertices of upward and downward triangles, placed on a honeycomb lattice,
are connected by bonds.
In the present model, additional sites are inserted in the middle of bonds connecting the triangles.
Therefore, we call this series of lattices decorated star lattices.

\begin{figure}[b]
\begin{center}
\includegraphics[width= 0.98\linewidth]{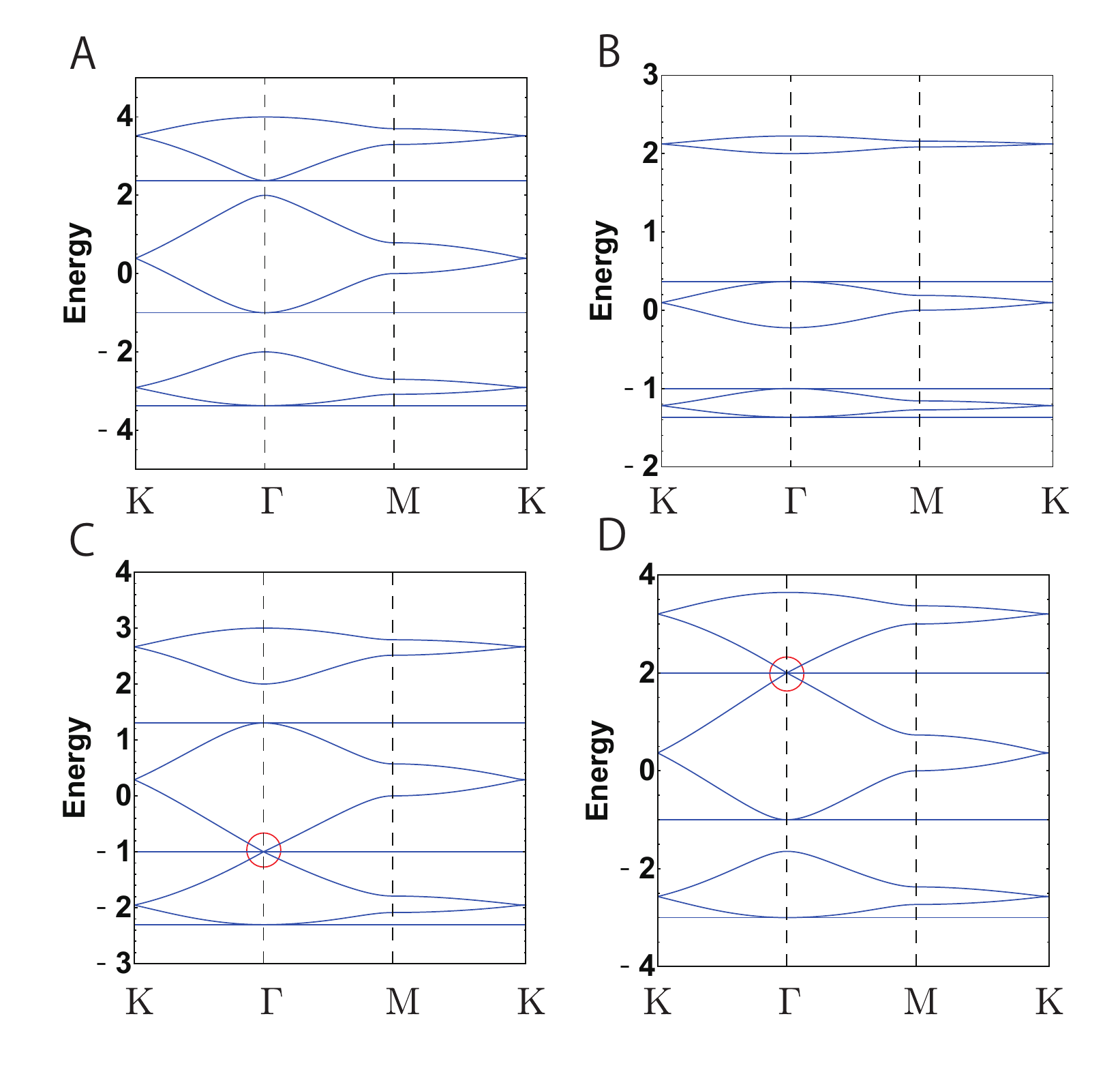}
\caption{The band structures for $t=1$ and 
A: $t^\prime = 2$, B: $t^\prime = 0.5$, C: $t^\prime = \sqrt{\frac{3}{2}}$, and D: $t^\prime = \sqrt{3}$.
The high-symmetry points in momentum space are $\Gamma = (0,0)$ K$=(\frac{4\pi}{3},0)$,
and M$=(\pi, \frac{\pi}{\sqrt{3}})$.}
\label{Fig2},
\end{center}
\end{figure}
Typical band structures for $t < t^\prime$ and
$t > t^\prime$ are shown in Fig.~\ref{Fig2} A and ~\ref{Fig2} B, respectively. 
It should be noted that
$t^\prime$ originates from the transfer integral between 
$\pi$ orbitals on the NN carbon atoms, 
while  $t$ from that on the next-NN carbon atoms~\cite{Fujii2018_2}.
Thus it is physically natural to set $t <  t^\prime$ rather than $t > t^\prime$ in the real material.
According to the first-principles calculations~\cite{Fujii2018_2}, $t$ is estimated to be 83 meV,
while $t^\prime$ can be tuned by the relative angle between the hexagonal planes of neighboring C$_6$ rings
(see Sec.~\ref{sec:tpbp} for details).

For any $t$ and $t^\prime$,  
there exist three flat bands, whose energies are 
$\varepsilon_{\mathrm{flat}}^{(1)} = -t$, 
$\varepsilon_{\mathrm{flat}}^{(2)} =-\frac{1}{2} \left(t + \sqrt{t^2 + 8t^{\prime 2}} \right)$,
and $\varepsilon_{\mathrm{flat}}^{(3)} =\frac{1}{2} \left(-t  + \sqrt{t^2 + 8t^{\prime 2}} \right)$.
In Appendix~\ref{sec:mo}, we explain how to obtain three flat-band energies;
the method is applicable to any length of chains. 
The existence of the flat bands means that the rank of the Hamiltonian matrix subtracted by the flat-band energy is reduced. 
To be concrete, consider the matrix $\bar{\mathcal{H}}^{(j)}_{\bm{k}} = \mathcal{H}_{\bm{k}} -\varepsilon_{\mathrm{flat}}^{(j)} E_9$ $(j=1,2,3)$.
Then, one can show that $\bar{\mathcal{H}}^{(j)}_{\bm{k}}$ can be rewritten by eight vectors, $\phi^{(j)}_{p,\bm{k}}$ ($p=1, \cdots 8$),
and a $8 \times 8$ matrix $h^{(j)}_{\bm{k}}$, as 
\begin{eqnarray}
\bar{\mathcal{H}}^{(j)}_{\bm{k}} =& \sum_{p,p^\prime = 1}^{8} \phi^{(j)}_{p,\bm{k}}  \left[h^{(j)}_{\bm{k}} \right]_{p p^\prime} \phi^{(j) \dagger}_{p^\prime,\bm{k}} \nonumber \\
=& \Phi^{(j)}_{\bm{k}} h^{(j)}_{\bm{k}} \Phi^{(j) \dagger}_{\bm{k}},  \label{eq:ham_MOrep}
\end{eqnarray}
with 
\begin{eqnarray}
\Phi^{(j)}_{\bm{k}} = \left[
\begin{array}{c}
\phi^{(j)}_{1,\bm{k}} \\
\phi^{(j)}_{2,\bm{k}} \\
\phi^{(j)}_{3,\bm{k}} \\
\phi^{(j)}_{4,\bm{k}} \\
\phi^{(j)}_{5,\bm{k}} \\
\phi^{(j)}_{6,\bm{k}} \\
\phi^{(j)}_{7,\bm{k}} \\
\phi^{(j)}_{8,\bm{k}} \\
\end{array}
\right]. \label{eq:projection}
\end{eqnarray}
Notice that each $\phi^{(j)}_{p,\bm{k}}$ is a nine-component row vector, and thus  $\Phi^{(j)}_{\bm{k}} $ is written as 
a $8 \times 9$ matrix (see Appendix~\ref{sec:mo} for details). 
Equation (\ref{eq:ham_MOrep}) manifests that the kernel of $\Phi^{(j)\dagger}_{\bm{k}}$
becomes the zero-energy eigenmode of $\bar{\mathcal{H}}^{(j)}_{\bm{k}}$ at each momentum $\bm{k}$. 
Equivalently, it is the eigenmode of $\mathcal{H}^{(j)}_{\bm{k}}$ with the eigenenregy $\varepsilon_{\mathrm{flat}}^{(j)}$, which is nothing but a flat band. 
The explicit forms of $\phi^{(j)}_{p,\bm{k}}$ and $h^{(j)}_{\bm{k}}$ in the present model
are given in Appendix~\ref{sec:mo}.
Since the real-space picture of $\phi^{(j)}_{p,\bm{k}}$,
obtained by the Fourier transformation, is composed of 
the superposition of the small number of sites, we refer to these real-space objects as ``MOs".
It should be remarked that these MOs are different from the ``true" MOs from the viewpoint of chemistry.  

Looking at the dispersive bands, one finds the Dirac cones at K and K$^\prime$ points, 
originating from the kagome network we describe in the next subsection.
Besides the Dirac cones at K and K$^{\prime}$ points, 
one also finds the triple band touching at $\Gamma$ point for a specific choice of $t^\prime$,
i.e., $t^\prime  = \sqrt{ \frac{3}{2} } t $ (Fig.~\ref{Fig2} C) and $t^\prime  = \sqrt{3} t$ (Fig.~\ref{Fig2} D). 
This is an example of a single Weyl fermion in lattice models crossed by a non-propagating mode, 
which was first argued by Dagotto, Fradkin, and Moreo~\cite{Dagotto1986}.
As is discussed in Ref.~\onlinecite{Dagotto1986}, this type of the band structure 
does not contradict the Nielsen-Ninomiya theorem on the fermion doubling in lattice models~\cite{Nielsen1981}, and
is indeed seen in various lattice models such as 
a Lieb lattice~\cite{Lieb1989},
a superhoneycomb lattice~\cite{Shima1993}, and a breathing kagome lattice 
with the upward triangles having opposite-sign hoppings to those on the downward ones~\cite{Essafi2017}. 

\begin{figure}[b]
\begin{center}
\includegraphics[width= .98\linewidth]{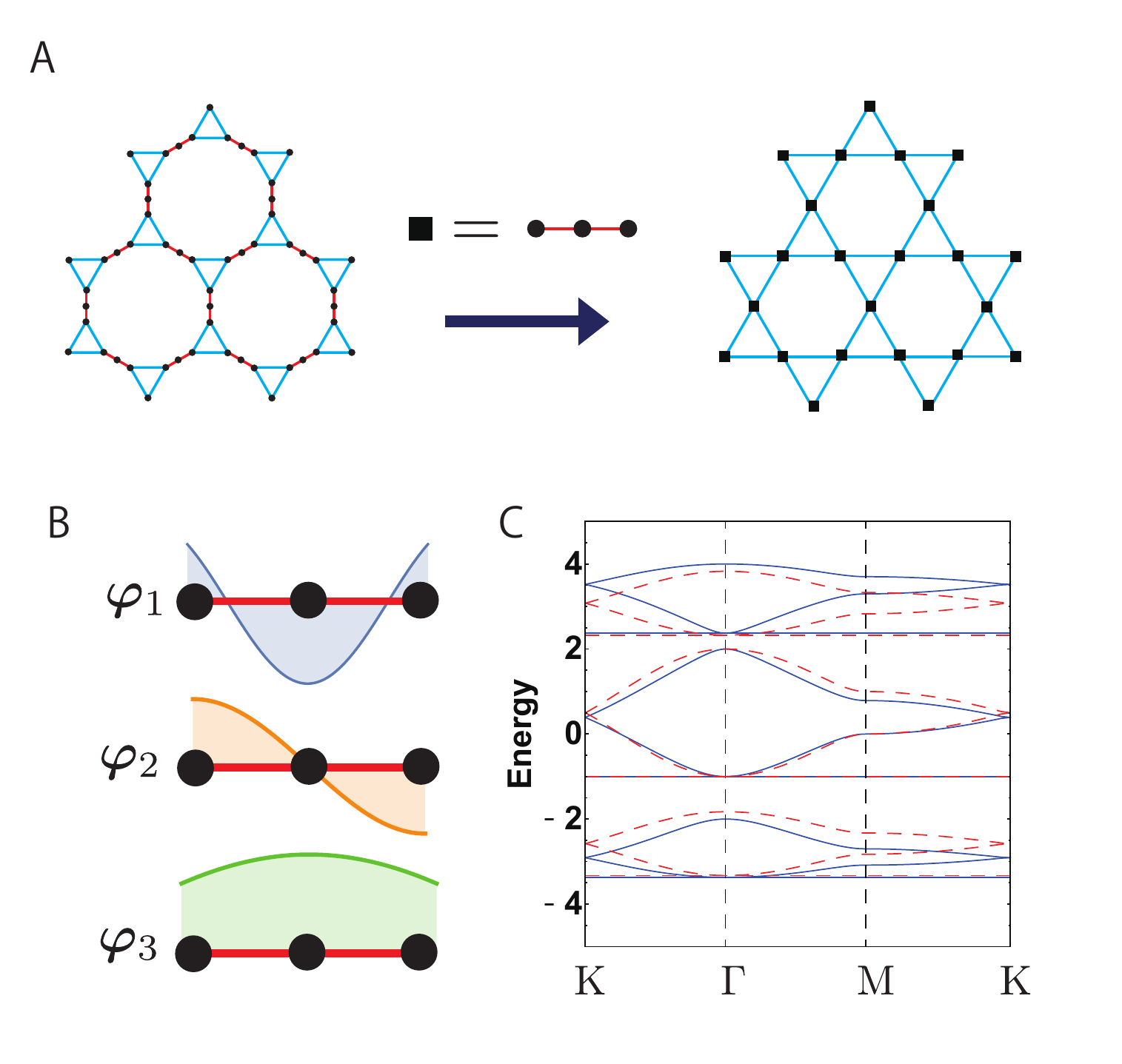}
\caption{
A: Schematic picture of the reduction of three sites connected by $t^\prime$-bonds to a single site,
and the resulting kagome lattice. 
B: Schematic picture of the real-space distribution of the three-site wave functions $\bm{\varphi}_1$-$\bm{\varphi}_3$.
C: The band structure for the Hamiltonian of Eq. (\ref{eq:ham}) 
with $(t,t^\prime) = (1,2) $
(blue solid lines) and the approximate band structure (red dashed lines). }
\label{Fig3}
\end{center}
\end{figure}
\subsection{Kagome network for polymerized triptycene containing phenyl \label{sec:kagomeapp}}
In Sec.~\ref{sec:kagomenet}, we mentioned that the kagome network in this material is obtained 
by regarding the chains of C$_6$ rings at the edges of the honeycomb lattice as a single point. 
Here, we employ the similar argument in the tight-binding model, 
in order to understand the band structures in terms of the kagome network. 
Looking at the lattice structure of Fig.~\ref{Fig1},
one can find that, if one regards three sites connected by the hopping of $t^\prime$ 
as one ``site", then one obtains the kagome lattice
(see Fig.~\ref{Fig3} A for the schematic picture). 
Since there are three degrees of freedom on each ``site", 
the band structure is approximated by that for the three-orbital kagome model. 

To formulate the above argument, consider the three sites connected by the bonds with the hopping $t^\prime$,
e.g., $(\bm{R},1)$, $(\bm{R},4)$, and $(\bm{R},7)$, where $\bm{R}$ denotes the position of the unit cell.
Then, consider the three-site Hamiltonian 
\begin{eqnarray}
h_{\rm 3-sites} = t^\prime \left( 
c^\dagger_{\bm{R},1} c_{\bm{R},4}  + c^\dagger_{\bm{R},4} c_{\bm{R},7}
\right) +(\mathrm{h.c.}). 
\end{eqnarray}
The eigenvalues and the eigenstates of $h_{\rm 3-sites}$, 
$\varepsilon_\xi$ and $a^\xi= \bm{\varphi}_{\xi} \cdot (c_{\bm{R},1} , c_{\bm{R},4}, c_{\bm{R},7} )^{\mathrm{T}}$ ($\xi=1,2,3$), 
respectively, are easily obtained as 
\begin{subequations}
\begin{eqnarray}
\varepsilon_{1} = -\sqrt{2} t^\prime, 
\end{eqnarray}
\begin{eqnarray}
\bm{\varphi}_1  = \frac{1}{2} \left(1,  - \sqrt{2} ,1 \right),
\end{eqnarray}  
\end{subequations}
\begin{subequations}
\begin{eqnarray}
\varepsilon_{2} = 0,
\end{eqnarray}
\begin{eqnarray}
\bm{\varphi}_2 = \frac{1}{\sqrt{2}} \left(-1,0,1 \right),
\end{eqnarray}  
\end{subequations}
\begin{subequations}
and
\begin{eqnarray}
\varepsilon_{3} = \sqrt{2} t^\prime,  
\end{eqnarray}
\begin{eqnarray}
\bm{\varphi_3} = \frac{1}{2} \left(1,  \sqrt{2} ,1  \right). 
\end{eqnarray}  
\end{subequations}
Schematic figures of $\bm{\varphi}_1$-$\bm{\varphi}_3$ are shown in Fig.~\ref{Fig3} B.
Similar three-site wave functions can be constructed 
for $[(\bm{R} , 2 ), (\bm{R},5) , (\bm{R},8)]$ and 
$[(\bm{R} ,3) , (\bm{R} 6), (\bm{R}, 9)]$.
Then, by 
using the basis $a^\xi_i$
with $i$ being the site on a kagome lattice,
the Hamiltonian can be then rewritten 
in a form of the three-orbital kagome model,
with the on-site potential $\varepsilon_{\xi}$ and the hopping integral
\begin{eqnarray}
T_{i,j}^{\xi, \eta} = \bra{\varphi_i^\xi}H^{\prime} \ket{\varphi_j^\eta},  \label{eq:capitalT}
\end{eqnarray} 
where $j$ is the NN site of $i$, 
$\ket{\varphi_i^\xi}$ denotes the wave fucntion corresponding to $a^\xi_i$, 
and $H^{\prime}$ denotes the part of the Hamiltonian $H^{\mathrm{P}}$ of Eq. (\ref{eq:ham_tp})
depends only on $t$ (i.e., blue triangles in Fig.~\ref{Fig1}). 

The above transformation of the Hamiltonian does not include any approximations, 
because we just change the basis from $c$ to $a$.
We now approximate the band structure as follows:
If $t \ll t^{\prime}$, the difference of the on-site energy $\varepsilon_\xi$ 
between different $\xi$s is much larger than $T_{i,j}^{\xi, \eta}$, thus 
we can neglect the hoppings between $T_{i,j}^{\xi, \eta}$ with $\xi \neq \eta$. 
If we do so, the band structure thus obtained is equal to that of three NN kagome bands
with different on-site energies. 
The corresponding NN hoppings are
\begin{eqnarray}
t_1 = T_{i,j}^{1,1} = \frac{t}{4},  
\end{eqnarray}
\begin{eqnarray}
t_2 = T_{i,j}^{2,2} = \frac{t}{2},  
\end{eqnarray}
and 
\begin{eqnarray}
t_3 = T_{i,j}^{3,3} = \frac{t}{4}.  
\end{eqnarray}
In Fig.~\ref{Fig3} C, we show the true band structure (blue solid lines) 
and the dispersion obtained by the present approximation (red dashed lines), with $(t,t^\prime) = (1,2)$.
They show a good agreement, besides the band width of the top and bottom kagome bands.  

\subsection{Polymerized triptycene containing biphenyl \label{sec:tpbp}}
\begin{figure}[t]
\begin{center}
\includegraphics[width= 0.98\linewidth]{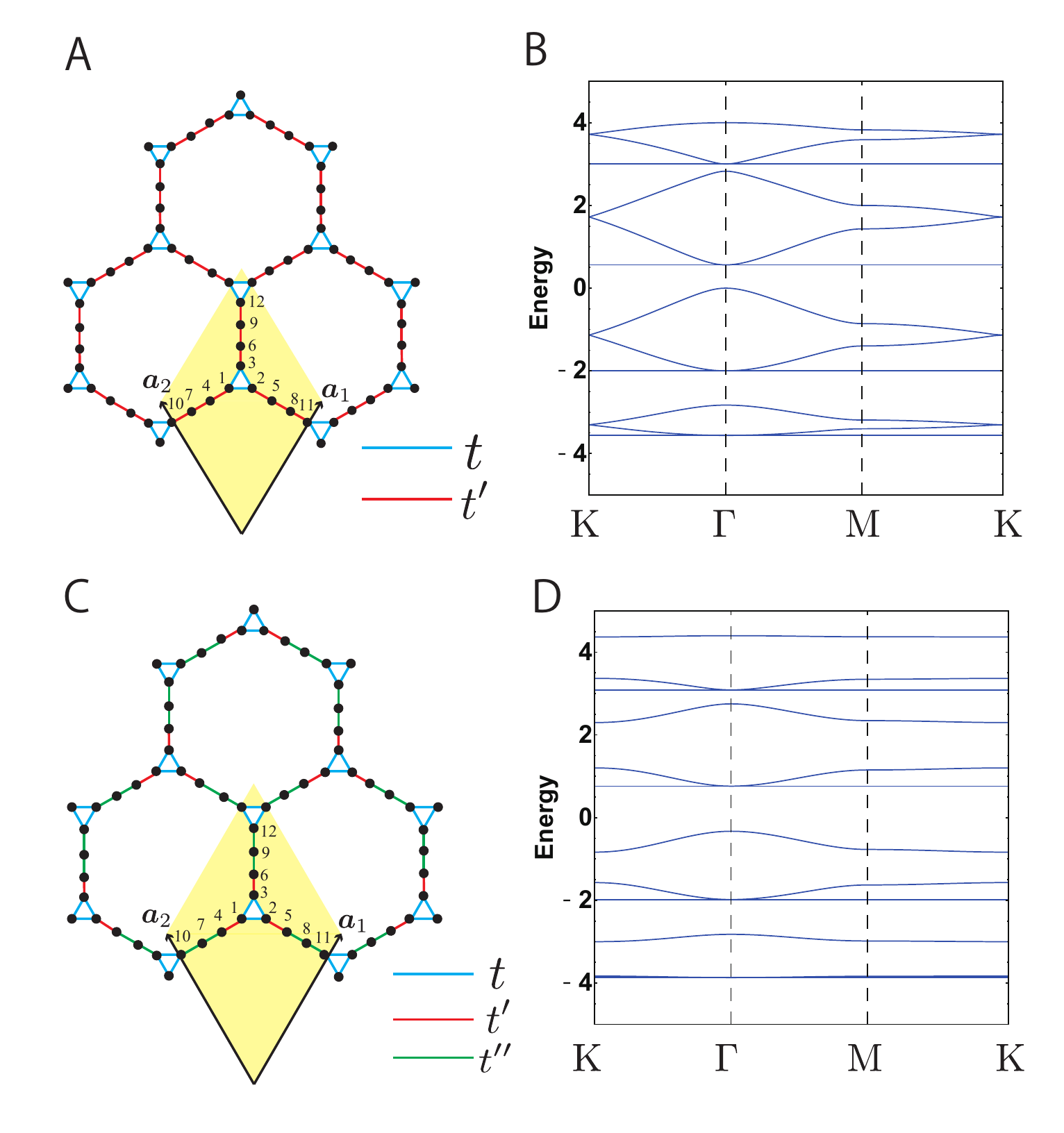}
\caption{A: The tight-binding model of Eq. (\ref{eq:hamBP}).
The primitive vectors and the sublattices are indicated in the figure. 
B: The band structure for $t=1$, $t^\prime = 2$. 
C: The tight-binding model of Eq. (\ref{eq:hamBP2}).
D: The band structure for $(t,t^\prime,t^{\prime \prime}) = (1,3,1.5)$}
\label{Fig4}
\end{center}
\end{figure}
For a series of polymerized triptycene, one can consider the materials in which the number of phenyls connecting the triptycene molecules is more than one~\cite{Fujii2018_2}. 
If there are two phenyls, the corresponding tight-binding model can be obtained by increasing the number of sites 
inserted in the bonds connecting the triangles (Fig.~\ref{Fig4} A),
and the Hamiltonian can be written as 
\begin{equation}
H^{\mathrm{BP}}= \bm{c}_{\bm{k}}^\dagger \mathcal{H}^{\mathrm{BP}}_{\bm{k}} \bm{c}_{\bm{k}}, \label{eq:hamBP}
\end{equation}
with 
\begin{eqnarray}
\mathcal{H}^{\mathrm{BP}}_{\bm{k}} = 
\left[
\begin{array}{cccc}
t h^{\bigtriangleup} & t^\prime E_3 & 0 & 0 \\
t^\prime E_3 & 0 & t^\prime E_3 & 0  \\
0 & t^\prime E_3 & 0 & t^\prime E_3 \\
0 & 0& t^{\prime} E_3 & t h^{\bigtriangledown}_{\bm{k}}  \\
\end{array}
\right], 
\end{eqnarray}
 and $\bm{c}_{\bm{k}} = \left[ c_{\bm{k},1},  \cdots, c_{\bm{k},12}  \right]^{\rm T}$.
A typical band structure for $t < t^\prime$ is shown in Fig.~\ref{Fig4} B.
We see that the number of flat bands is four;
this implies that, in general, the number of flat bands in this series of 
tight-binding models with $q$ sites between triangles is equal to $q + 2$. 
This can be proved by explicitly constructing the flat bands; see Appnedix~\ref{sec:mo} for details.

One of the interesting consequences 
of increasing the number of inter-triptycene phenyls 
is that one can tune the NN hopping between 
$\pi$ orbitals on the phenyls by changing 
the relative angle of the hexagonal planes of C$_6$ rings.
For a particular configuration of C$_6$ rings, one obtains the tight-binding Hamiltonian~\cite{Fujii2018_2}: 
\begin{equation}
\tilde{H}^{\mathrm{BP}}= \bm{c}_{\bm{k}}^\dagger \tilde{\mathcal{H}}^{\mathrm{BP}}_{\bm{k}} \bm{c}_{\bm{k}}, \label{eq:hamBP2}
\end{equation}
with 
\begin{eqnarray}
\tilde{\mathcal{H}}^{\mathrm{BP}}_{\bm{k}} = 
\left[
\begin{array}{cccc}
t h^{\bigtriangleup} & t^\prime E_3 & 0 & 0 \\
t^\prime E_3 & 0 & t^{\prime \prime} E_3 & 0  \\
0 & t^{\prime  \prime} E_3 & 0 & t^{\prime  \prime} E_3 \\
0 & 0& t^{\prime \prime} E_3 & t h^{\bigtriangledown}_{\bm{k}}  \\
\end{array}
\right].  \label{eq:ham_bp}
\end{eqnarray}
The schematic figure of the model of Eq. (\ref{eq:hamBP2}) is shown in Fig.~\ref{Fig4} C.
Here a new parameter $t^{\prime \prime} (\neq t^\prime) $ is introduced, 
reflecting the difference of the relative angles of the hexagonal planes.
A typical band structure for $t <t^{\prime \prime}< t^\prime$ is shown in Fig.~\ref{Fig4} D.
Although there still exist four flat bands, the massless Dirac dispersion
at K point observed in Fig.~\ref{Fig4} B is now gapped out,
which is reminiscent of the breathing kagome lattice~\cite{Ezawa2018,Xu2017,Kunst2018,Hatsugai2011}.
\begin{figure}[b]
\begin{center}
\includegraphics[width= 0.95\linewidth]{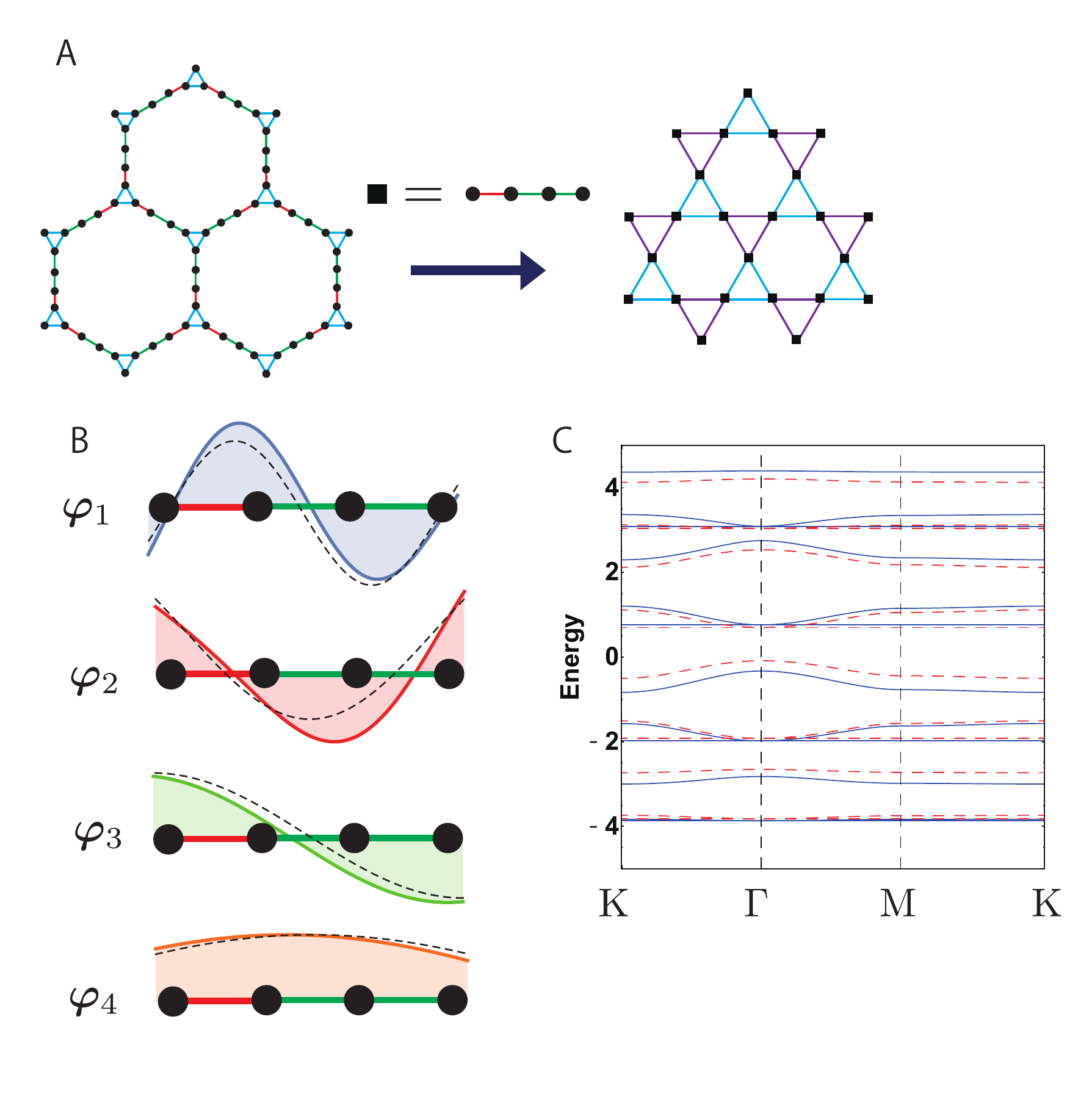}
\caption{A: Schematic picture of the reduction of four sites connected by 
$t^\prime$- or $t^{\prime \prime}$-bonds to a single site,
and the resulting breathing kagome lattice. 
B: Schematic picture of the real-space distribution of the four-site wavefunctions $\bm{\varphi}_1$-$\bm{\varphi}_4$.
Dashed lines denote the case for $t^\prime=t^{\prime \prime}$, where the inversion symmetry is restored.
C: The band structure for the Hamiltonian of Eq. (\ref{eq:hamBP2}) 
with $(t,t^\prime,t^{\prime \prime}) = (1,3,1.5)$
(blue solid lines) and the approximate band structure (red dashed lines).  }
\label{Fig:appkagome_2}
\end{center}
\end{figure}

\subsection{Kagome network for polymerized triptycene containing biphenyl \label{sec:tpbp_kagome}}
Similar to the case of $H^{\mathrm{P}}$, 
we can apply the approximate kagome description to $\tilde{H}^{\mathrm{BP}}$ (Fig.~\ref{Fig:appkagome_2} A). 
In this case, we reduce four sites connected by $t^\prime$ and $t^{\prime \prime}$
to one site, 
which allows us to employ the four-orbital kagome description (Fig.~\ref{Fig:appkagome_2} B).
Interestingly, due to the imbalance between $t^\prime$ or $t^{\prime \prime}$,
the four-site wave function $\bm{\varphi}_\xi$ becomes inversion asymmetric.
Consequently, the hoppings for $\bm{\varphi}_\xi$ acquire the breathing structure, 
namely,
$T_{i,j | \langle i,j \rangle \in \bigtriangleup}^{\xi, \xi} \neq T_{i,j | \langle i,j \rangle \in \bigtriangledown}^{\xi, \xi}$,
which, as pointed out in the above, becomes the source of the massive Dirac dispersion at K point (Fig.~\ref{Fig:appkagome_2} C). 
\begin{figure*}[t]
\begin{center}
\includegraphics[width= 0.98\linewidth]{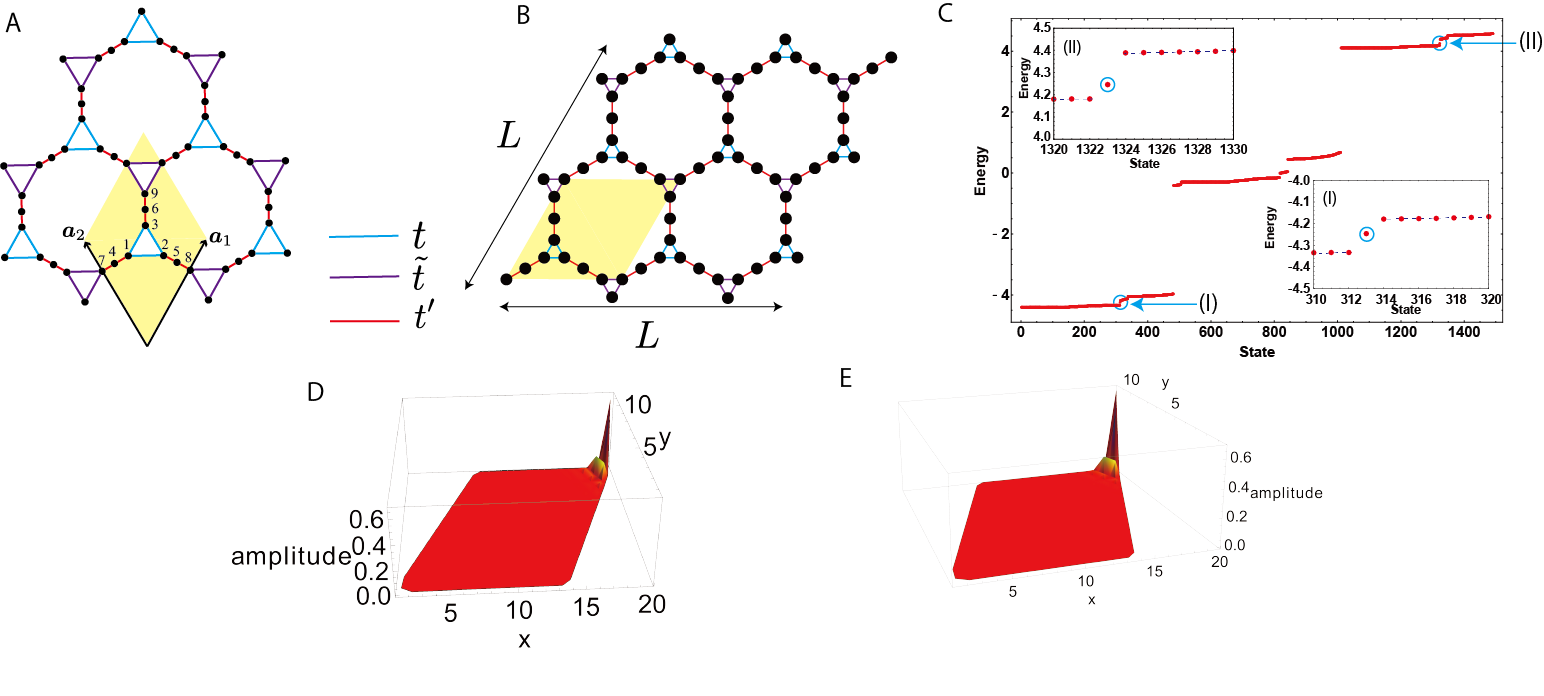}
\caption{A: The tight-binding model for Eq. (\ref{eq:tri_bk}). 
B: Schematic figure of the finite system used to investigate the corner state. 
C: The energy spectrum for the finite system with $(t,\tilde{t},t^\prime) = (0.5,0.1,3)$ and $L=12$
(the number of sites is 1491).
Two in-gap states are indicated by light-blue circles;
here (I) is 313th state and (II) is 1323th state.
Insets are energy levels near the in-gap states.
The amplitude of the wave functions in the real space
for D: the state (I) and E: the state (II).}
\label{Fig:corner1}
\end{center}
\end{figure*}

\section{Possibility of higher-order topological phase \label{sec:hoti}} 
Having the approximate kagome picture at hand, 
we now propose how to realize the HOTI phase in this material. 
Previous studies have revealed that
the HOTI phase on a kagome lattice is obtained by introducing a ``breathing" structure~\cite{Ezawa2018,Kunst2018,Xu2017,Araki2019},
i.e., the imbalance of hopping integrals between upward triangles and downward triangles. 
This is essentially a two-dimensional analog of the Su-Schriffer-Heeger model~\cite{Su1979},
where the gapless end state appears associated with the non-trivial winding number.

Our strategy is to mimic the breathing kagome structure in the present materials. 
In this paper, we examine two ways to realize the HOTI in polymerized triptycene as follows:
\begin{itemize}
\item[(i)] We consider a material where triptycene molecules placed on the downward triangles in Fig.~\ref{Fig1}
are replaced with some other molecules 
(see Fig.~\ref{Fig:corner1} A). 
Note that we neglect the difference of on-site energies between triptycene molecules and the newly-introduced molecules.
In the tight-binding description, the hoppings on downward triangles are changed from $t$ to $\tilde{t} \neq t$. 
In this case, the Hamiltonian matrix given as follows:
\begin{equation}
\mathcal{H}^{\rm (i)}_{\bm{k}} = \left[
\begin{array}{ccc}
t h^\bigtriangleup &t^\prime E_3  &0 \\
t^\prime E_3 & 0& t^\prime E_3 \\
0 &  t^\prime E_3 & \tilde{t} h^\bigtriangledown_{\bm{k}} \\
\end{array}
\right]. \label{eq:tri_bk}
\end{equation}
One can easily see that the approximate network of such a material 
corresponds to the breathing kagome lattice, 
by using the same method as in \ref{sec:kagomeapp}.

\item[(ii)] As we have already seen in the previous section, 
the polymerized triptycene containing biphenyl with $t^\prime \neq t^{\prime \prime}$ 
(Fig.~\ref{Fig4} C)
is a candidate of the HOTI, 
since its approximate network corresponds to the breathing kagome structure. 
\end{itemize}

\subsection{Corner states under the open boundary conditions}
\begin{figure*}[t]
\begin{center}
\includegraphics[width=\linewidth]{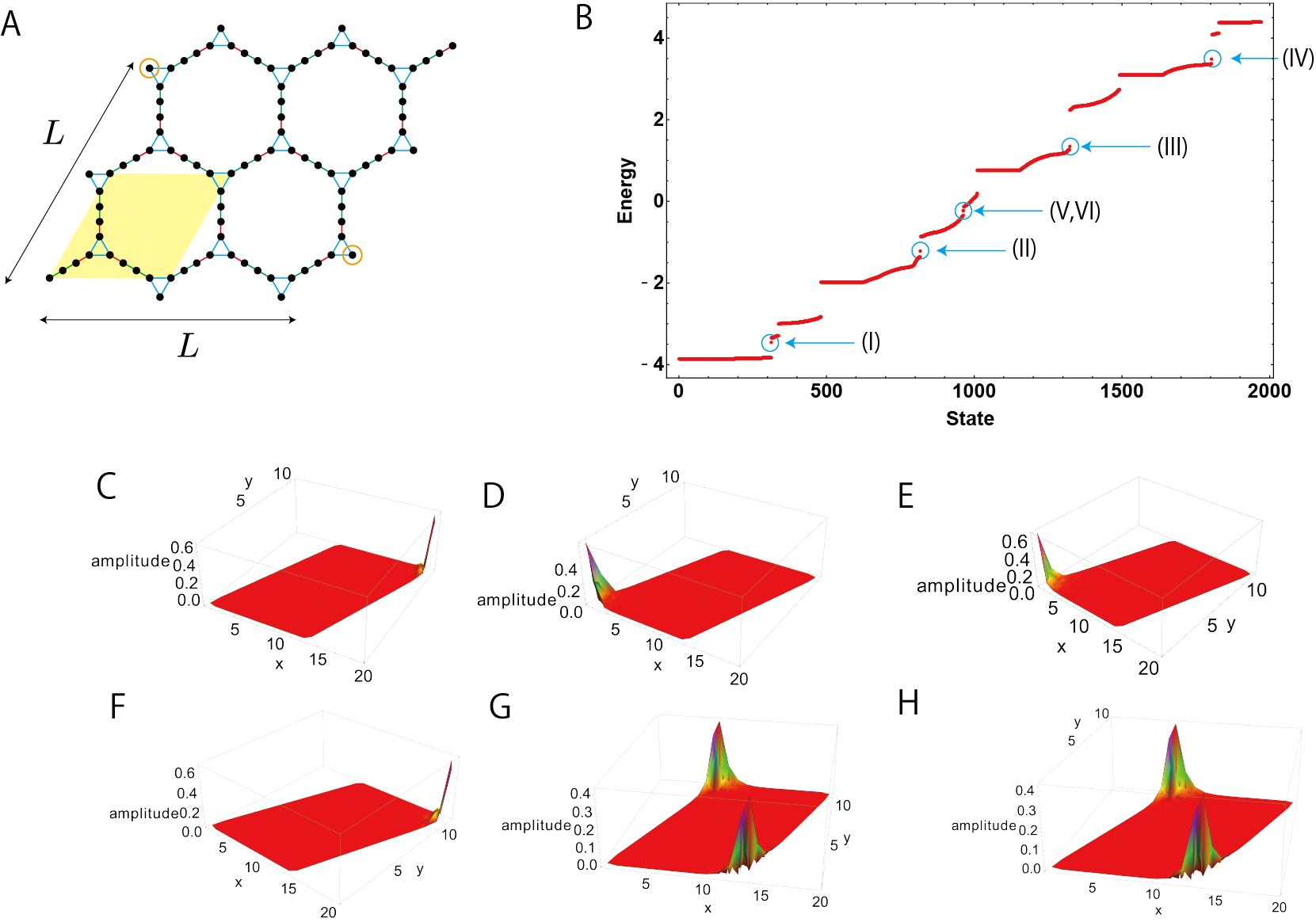}
\caption{ 
A: Schematic figure of the finite system used to investigate the corner state in polymerized triptycene containing biphenyl. 
Orange circles denote the sites at bottom-left and top-right corners. 
B: The energy spectrum for the finite system with $(t,t^{\prime},t^{\prime \prime}) = (1,3,1.5)$ and $L=12$
(the number of sites is 1972).
Six in-gap states are indicated by light-blue circles;
here (I) is 313th state, (II) is 818th state, (III) is 1323th state, (IV) is 1804th state, 
and (V) and (VI), which are degenerate, are 963th and 964th states.
C-H: The amplitude of the wave functions in the real space.
C, D, E, F, G, and H correspond to (I), (II), (III), (IV) (V), and (VI), respectively. }
\label{Fig:corner2}
\end{center}
\end{figure*}

To demonstrate the realization of the HOTI, we first examine
the existence of the corner states.
We diagonalize the Hamiltonian on the finite systems in a rhombus geometry,
shown in Fig.~\ref{Fig:corner1} B and Fig.~\ref{Fig:corner2} A for the cases (i) and (ii), respectively. 
Note that the edges and corners are chosen such that the blue or purple triangles are not cut off, since, from the materials point of view, they represent the triptycene molecules.

Let us first consider the case (i).
In Fig.~\ref{Fig:corner1} C, the energy eigenvalues are plotted. 
We see two in-gap states, whose energies are located between two dispersive bands of the breathing kagome band.
Figures~\ref{Fig:corner1} D and \ref{Fig:corner1} E represent the amplitudes 
of the in-gap eigenstates in the real space. 
They are sharply localized at the top-right corner of the rhombus sample, 
indicating that the HOTI phase is realized in this material. 
It is naively expected that each breathing kagome band has one corner state between the two dispersive bands~\cite{Ezawa2018,Kunst2018,Xu2017},
meaning that there are three in-gap states.
In the present case, however, the middle kagome band does not host the in-gap state,
indicating that the corner state is buried in the bulk or edge state.

Figure~\ref{Fig:corner2} illustrates the results for the case (ii).
We see six in-gap states (Fig.~\ref{Fig:corner2} B); among them, the states (I)-(IV) are located between the two dispersive bands of the breathing kagome bands, while the states (V)-(VI) are located between the different breathing kagome bands.

In Figs.~\ref{Fig:corner2} C-H, we depict the wave functions of the in-gap states (I)-(VI).
Among them, the states (I)-(IV) are localized at bottom-left or top-right corners, as is the case with the breathing kagome lattice~\cite{Xu2017,Kunst2018}.
In contrast, for the states (V) and (VI) are localized at the bottom-right and the top-left corners, which cannot be accounted for the breathing kagome picture.
The emergence of (V) and (VI) may be attributed to the fact that the sites at bottom-right and the top-left corners (the orange circles in Fig.~\ref{Fig:corner2} A) do not belong to the four-site chain of which the breathing kagome network is composed. 
Note that these sites cannot be removed since, as mentioned before, they are parts of the triptycene molecule.

It is interesting to find that, for the states (I)-(IV), the position of the site having the maximum amplitude differs from state to state.
More precisely, the states (I) and (IV) are the top-right corner state, whereas the states (II) and (III) are the bottom-left corner state.
This difference originates from the fact that the breathing pattern for 
the first and the fourth kagome bands is opposite to that for the second and the third kagome band.
To be more specific, 
\begin{eqnarray}
|T_{i,j | \langle i,j \rangle \in \bigtriangleup}^{1,1} | &>& | T_{i,j | \langle i,j \rangle \in \bigtriangledown}^{1,1} | \nonumber \\
|T_{i,j | \langle i,j \rangle \in \bigtriangleup}^{4,4} | &>& | T_{i,j | \langle i,j \rangle \in \bigtriangledown}^{4,4} |, \label{eq:relation1}
\end{eqnarray}
and 
\begin{eqnarray}
|T_{i,j | \langle i,j \rangle \in \bigtriangleup}^{2,2} | &<& | T_{i,j | \langle i,j \rangle \in \bigtriangledown}^{2,2} | \nonumber \\
|T_{i,j | \langle i,j \rangle \in \bigtriangleup}^{3,3} | &<& | T_{i,j | \langle i,j \rangle \in \bigtriangledown}^{3,3} |, \label{eq:relation2}
\end{eqnarray}
are satisfied in the present choice of parameters, $(t,t^{\prime},t^{\prime \prime})$.
It follows from Eq. (\ref{eq:capitalT}) that the amplitude of $T_{i,j | \langle i,j \rangle \in \bigtriangleup}^{\xi, \xi}$ is determined 
by the real-space distribution of $\bm{\varphi}_\xi$.
Hence, the above relations of Eqs. (\ref{eq:relation1}) and (\ref{eq:relation2}) indicate that 
$\bm{\varphi}_1$ and $\bm{\varphi}_4$ have a large amplitude at the left edge in Fig.~\ref{Fig:appkagome_2} B,
while $\bm{\varphi}_2$ and $\bm{\varphi}_3$ have a small amplitude. 

\subsection{$\mathbb{Z}_3$ Berry phase}
\begin{figure*}[t]
\begin{center}
\includegraphics[width= 0.85\linewidth]{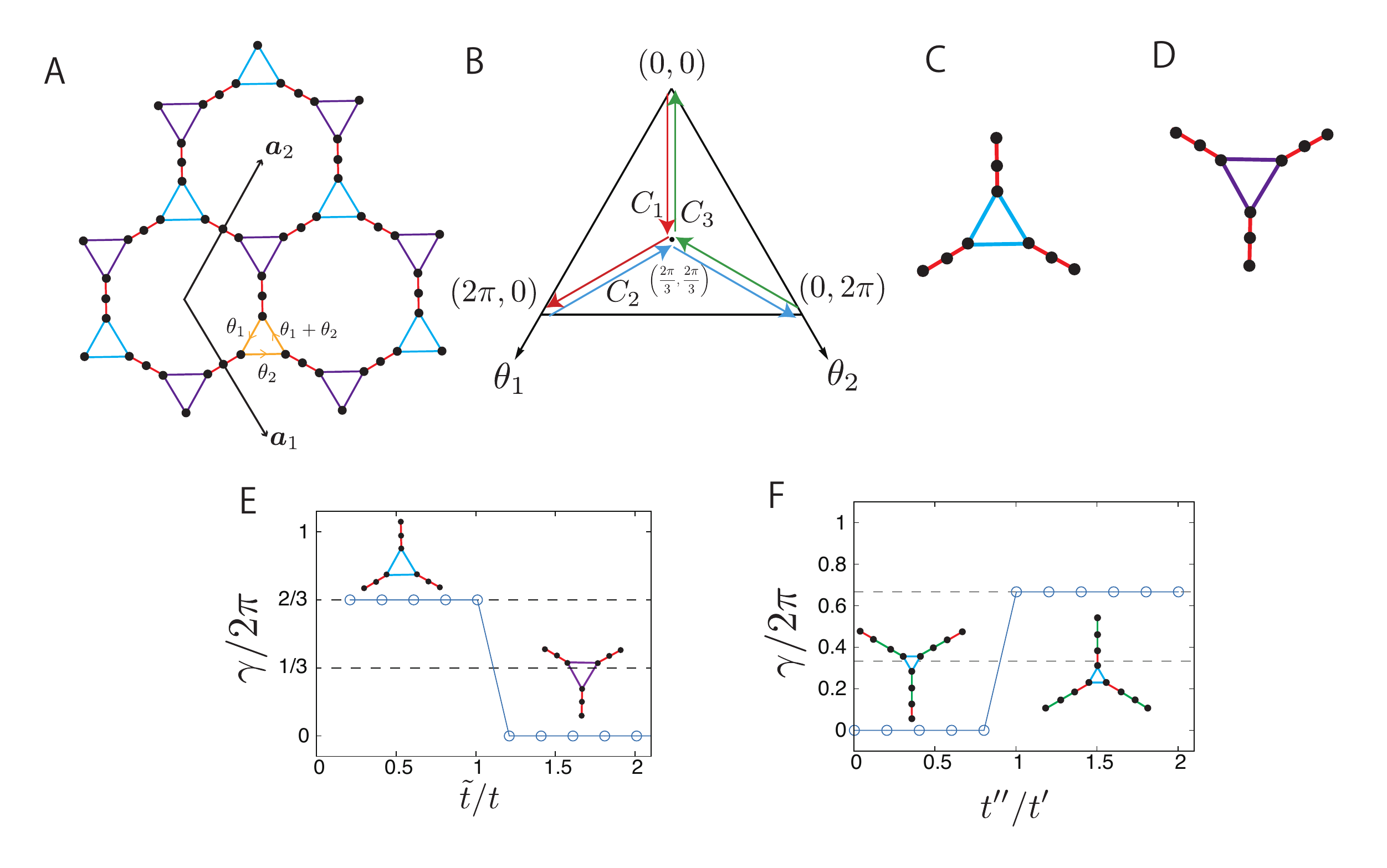}
\caption{A: Schematic figure for the twist introduced in Eq. (\ref{eq:twistedhamiltonian}). 
B: The paths of the contour integral in Eq. (\ref{eq:berry}) in the parameter space.
C and D: The minimal cluster for the Hamiltonian of Eq. (\ref{eq:tri_bk}).
C is obtained by setting $\tilde{t} = 0$, while D is obtained by setting $t=0$. 
E: $\tilde{t}/t$-dependence of the $\mathbb{Z}_3$ Berry phase 
for the model of Eq. (\ref{eq:tri_bk}) [i.e., case (i)]
with $2/9$-filling. 
The system size used for the computation is $12  \times 12 $ unit cells. 
F: $t^\prime/t^{\prime \prime}$-dependence of the $\mathbb{Z}_3$ Berry phase for the model of Eq. (\ref{eq:ham_bp})
[i.e., case (ii)] with $5/12$-filling. 
The system size used for the computation is $8 \times 8$ unit cells. 
The insets in E and F represent the minimal clusters to which the ground states are connected adiabatically.
 }
\label{Fig:Berry}
\end{center}
\end{figure*}
To demonstrate that the corner states we have shown in the previous subsection are indeed 
topologically protected, we calculate the topological invariant that characterizes the HOTI states, namely the 
$\mathbb{Z}_3$ Berry phase~\cite{Hatsugai2011,Kariyado2018,Kawarabayashi2019,Kudo2019,Araki2019_2}.
The $\mathbb{Z}_3$ Berry phase is defined with respect to the parameters $\bm{\Theta} = (\theta_1, \theta_2)$,
which is included in the twisted Hamiltonian, $H(\bm{\Theta})$.
To be more specific, we define $H(\bm{\Theta})$ as follows:
First, we pick up an upward triangle at the unit cell labeled by $\bm{R}_0$,
and decompose the Hamiltonian as $H = h_0 + (H-h_0)$,
where 
\begin{eqnarray}
h_0 = t\left ( c^\dagger_{\bm{R}_0, 1} c_{\bm{R}_0, 2}
+ c^\dagger_{\bm{R}_0, 2} c_{\bm{R}_0, 3 } 
 + c^\dagger_{\bm{R}_0, 3} c_{\bm{R}_0, 1}  \right)
+({\rm h.c.}). \nonumber \\
\end{eqnarray}
Then, we introduce the ``twist" of the Hamiltonian only at $h_0$ 
as $H(\bm{\Theta}) = h_0 (\bm{\Theta}) +  (H-h_0)$, 
where $\bm{\Theta} = (\theta_1,\theta_2)$ 
with $\theta_1 \in [0,2\pi]$, $\theta_2 \in [0,2\pi]$,
and 
\begin{eqnarray}
h_0(\bm{\Theta}) &=& te^{i \theta_2} c^\dagger_{\bm{R}_0, 1} c_{\bm{R}_0, 2}
 + te^{- i (\theta_1 + \theta_2) }c^\dagger_{\bm{R}_0, 2} c_{\bm{R}_0, 3 } \nonumber \\
 &+& te^{i \theta_1} c^\dagger_{\bm{R}_0, 3} c_{\bm{R}_0, 1} 
+ ({\rm h.c.}).
\label{eq:twistedhamiltonian} 
\end{eqnarray} 
See Fig.~\ref{Fig:Berry} A for the schematic figure of Eq. (\ref{eq:twistedhamiltonian}). 
Note that $H(\bm{\Theta})$ does not have a translational symmetry due to the local twist. 
Now, let $| \Phi_0 (\bm{\Theta}) \rangle  $ be the many-body ground state of $H(\bm{\Theta})$.
Then, the $\mathbb{Z}_3$ Berry phase is defined as a contour integral
of the Berry connection along the path $C_\eta$ with $\eta = 1,2,3$, shown in Fig.~\ref{Fig:Berry} B:  
\begin{eqnarray}
\gamma_{\eta} \equiv  -i \int_{C_\eta} d \bm{\Theta} \cdot \langle \Phi_0 (\bm{\Theta}) | \nabla_{\Theta} \Phi_0 (\bm{\Theta}) \rangle \hspace{0.5mm} ({\rm mod} \hspace{.5mm} 2\pi). \label{eq:berry}
\end{eqnarray}
The (fractional) quantization of $\gamma_{\eta}$ is enforced by $C_3$ symmetry of the Hamiltonian.
Namely, it is clear from Fig.~\ref{Fig:Berry} B that $\sum_{\eta=1}^3 \gamma_{\eta} \equiv 0 $ due to the cancellation of the paths,
while the $C_3$ symmetry leads to $\gamma_{1} \equiv \gamma_{2} \equiv \gamma_{3}$.
Combining these, we have $\gamma_{\eta} = n \cdot \frac{2 \pi}{3}$ with $n=0,1,2$.
In the following, we abbreviate $\gamma_{\eta}$ as $\gamma$.

The nontrivial Berry phase indicates that the ground state is adiabatically connected to 
the product state of the ``minimal clusters"~\cite{Hatsugai2011,Kariyado2018,Kawarabayashi2019,Kudo2019,Araki2019_2}.
Here, the term ``minimal means that the cluster cannot be decomposed 
into the smaller element under the restriction imposed by the symmetry,
without closing the energy gap.  
The minimal cluster can be obtain by switching off some 
hoppings in the original tight-binding Hamiltonian
[see Figs.~\ref{Fig:Berry} C and  \ref{Fig:Berry} D for the model in Eq. (\ref{eq:tri_bk})]. 
In general, the Berry phase takes nontrivial value when 
the minimal cluster contains the bonds where we have introduced the twist in $H(\bm{\Theta})$~\cite{Hatsugai2011,Kariyado2018,Kawarabayashi2019,Kudo2019,Araki2019_2}.

The decoupled cluster picture indicates that the corner states appear under the open boundary condition 
when the minimal clusters are cut off at the boundaries~\cite{Hatsugai2011,Araki2019_2}.
This picture provides us of a simple understanding of the bulk-boundary correspondence~\cite{Hatsugai1993} 
in the various symmetry-protected topological phases. 
Therefore, the quantized Berry phase serves as a topological invariant 
for the HOTI phase, and is associated with the emergence of the corner states under appropriate choices of boundaries. 

Turning to the present models, we plot $\gamma$ for the case (i) with $2/9$-filling and case (ii) with $5/12$-filling, in Figs.~\ref{Fig:Berry} E and \ref{Fig:Berry} F respectively,
as a function of the parameter which controls the breathing of the kagome bands. 
Note that we follow Ref.~\onlinecite{Hatsugai2006_2} for the method of the numerical computations of the Berry phase. 
We clearly see that the topological phase transition, associated with the change of the Berry phase, 
occurs at the point where the breathing structure is lost, i.e., $t= \tilde{t}$ for the case (i) and $t^{\prime} = t^{\prime \prime}$ for the case (ii).
At the transition point, the bulk band gap at the K point closes as we have seen in Sec.~\ref{sec:model}. 

\section{Discussions \label{sec:discussion}}
So far, we have demonstrated that the polymerized triptycene is a promising candidates 
for the HOTI in the solid-state systems, by investigating the tight-binding models for the spinless fermions. 
In this section, we discuss some perspectives beyond the tight-binding analysis.

Firstly, for the observation of the HOTI in the present system, robustness of the corner states against disorders 
is important because disorders exist inevitably in solid-state systems. 
In the present system, the HOTI phase is protected by the $C_3$ symmetry, while disorders break this symmetry.
Hence, it is not trivial whether or not the corner states are stable against the disorders. 
In fact, the previous study has revealed that the corner states in the breathing kagome model
is stable against the nonmagnetic impurities, as far as the bulk band gap remains to be opened~\cite{Araki2019}.
We therefore expect that the corner states in polymerized triptycene behave in the similar way when the disorders are introduced,
and thus can be observed experimentally even in the presence of weak disorders. 
Meanwhile, the corner states are sensitive to the shape of the boundaries. 
For instance, if we remove of the sites at left-bottom or right-top corners, 
the energy of the corner states will change and they will be buried in the bulk continuum. 
Thus, fabrication of the proper corner shape is important for experimental observation of the corner states.

Secondly, the effect of electron-electron correlation will also give rise to interesting properties. 
To be concrete, consider the effect of the Hubbard interaction.
Note that we need to restore the spin degrees of freedom to consider the Hubbard interaction.
The physical properties in the presence of the Hubbard interaction are crudely dependent on the Fermi level. 
If the Fermi level is right at the flat band, one expects that flat-band ferromagnetism occurs, as is inferred from the results of studies based on the spin-dependent density functional theory for related materials~\cite{Maruyama2016,Maruyama2017,Sorimachi2017}. 
In contrast, if the free-electron ground state is the HOTI phase, 
where the Fermi level is located at the corner-state energy, 
infinitesimally small on-site interaction is enough to 
gap out the charge excitation at the corner, i.e., double occupancy of the corner state costs the finite energy.
We can therefore expect that the present system will be a good candidate to search the higher-order topological Mott insulator (HOTMI),
which was proposed recently for the breathing kagome model~\cite{Kudo2019}. 

Finally, we address the experimental observation of the HOTI phase. 
To observe the HOTI phase, 
one has to tune the filling of electrons such that Fermi level is located 
in the middle of the band gap so that two out of three bands are completely filled for a certain breathing kagome band. 
Comparing our tight-binding model and the real material, 
the system is either full-filled,
if we regard our model as that for the highest occupied molecular orbital of C$_6$ ring,
or totally empty, for the lowest unoccupied molecular orbital.
So, the electron or hole doping is necessary, and this can be achieved either by chemical doping or by placing the sample on a substrate and applying the gate voltage~\cite{Maruyama2016}. 

\section{Summary \label{sec:summary}}
To summarized, we have investigated the characteristic 
band structures of a family of polymerized triptycene
by the tight-binding models on decorated star lattices. 
We have found that this class of tight-binding models is indeed a fertile ground to 
realize exotic band structures, such as flat bands, and HOTIs, which arise from the underlying kagome network. 
As for the flat bands, we have demonstrated their existence by using the MO-representation (see Appendix~\ref{sec:mo} for details).
As for the HOTIs, we have proposed two methods to realize the HOTIs, 
and have demonstrated the existence of the HOTI phase by 
directly showing the corner states on the finite system. 
Further, we have associated the corner states 
with the bulk topological invariant, namely the $\mathbb{Z}_3$ Berry phase.
We then conclude that these materials are
promising candidates of the HOTI in the solid-state systems.

Similar strategy to realize the HOTI phase can be applicable to 
various materials hosting the kagome-type network with tunable hopping parameters.
Metal-organic frameworks with kagome-type network~\cite{Yamada2016,Barreteau2017} will be suitable candidates.
Besides the kagome-type network, similar realization of the HOTI
was proposed in graphdiyne~\cite{Narita1998,Nomura2018}, 
where the hexagonal plaquettes are interconnected instead of triangles~\cite{Sheng2019,Lee2019}.
Further, a three-dimensional cousin of the present system has been also studied~\cite{Ben2009,Trewin2010,Fujii2019}, 
which will be a candidate for three-dimensional third-order topological insulator.  
We expect that the organic materials with polymerized structures will be a fertile ground to 
search the HOTIs. 

\textit{Note added.}
 Recently,
we became aware of the related work~\cite{Lee2019_2}
where the flat bands and corner states of the similar model are studied
in the context of 1T-TaS$_2$ in 
a charge density wave state~\cite{Park2019}.
We also found after completing the present manuscript that the decorated star lattice 
is formed in a newly-proposed carbon allotrope called cyclicgraphene~\cite{You2019}.

\acknowledgements
We thank T. Yoshida, H. Araki, K. Kudo, and Y. Fujii for fruitful discussions. 
This work is supported by the JSPS KAKENHI, Grant number JP17H06138 (T. M. and Y. H.), MEXT, Japan.

\begin{widetext}
\appendix
\section{Molecular-orbital representation: Exact treatment of the flat bands \label{sec:mo}}
In this Appendix, we clarify the origin of flat bands in the polymerized triptycene by using the framework which we call
the MO representation~\cite{Hatsugai2011,Hatsugai2015,Mizoguchi2019}. 

As we have shown in the main text, there are $q+2$ flat bands with different energies
for the system containing $q$ phenyls between triptycene molecules. 
Here, we first explain how to determine the flat-band energies.
Through this procedure, we find that there are indeed $q+2$ flat bands for generic $q$. 
Then, we show the explicit forms of the MOs which describe the Hamiltonian in the form of (\ref{eq:ham_MOrep}) for the case with $q=1$.
More precisely, we show that for each flat band, 
the Hamiltonian matrix $\mathcal{H}_{\bm{k}}^{\rm P} $ subtracted by the flat-band energy 
can be written by eight $\bm{k}$-dependent vectors
whose inverse Fourier transformations correspond to what we call MOs in the previous studies~\cite{Hatsugai2011,Mizoguchi2019}. 
This description of flat-band models with multiple flat bands was used in Ref.~\onlinecite{Hatsugai2015},
where the Weaire-Thorpe model~\cite{Weaire1971} was studied.

\subsection{Determination of flat-band energies}
Consider the Schr\"{o}dinger equation
\begin{eqnarray}
\mathcal{H}^{(q)}_{\bm{k}}  \bm{U}_{\bm{k}} = \varepsilon  \bm{U}_{\bm{k}}, \label{eq:sc}
\end{eqnarray}
where $\mathcal{H}^{(q)}_{\bm{k}}$ is the Hamiltonian in the form
\begin{eqnarray}
\mathcal{H}^{(q)}_{\bm{k}} = 
\left[
\begin{array}{cccccc}
t h^\bigtriangleup  & t_1E_3  & 0 & \cdots & \cdots& 0\\
t_1E_3 & 0 & t_2 E_3& \cdots &\cdots & 0 \\
0 & t_2 E_3&0& \cdots & \cdots& 0 \\
&  \vdots  & &\ddots& & \vdots  \\
0 & \cdots &\cdots &\cdots & 0 & t_{q+1} E_3\\
0 & \cdots &\cdots & \cdots& t_{q+1} E_3& th^\bigtriangledown_{\bm{k}} \\\end{array}
\right],
\end{eqnarray}
and 
$\bm{U}_{\bm{k}} = \left(u_{1,\bm{k}} , u_{2,\bm{k}}, \cdots,  u_{3(q+2),\bm{k}} \right)^{\rm T}$ is an eigenvector of $\mathcal{H}^{(q) }_{\bm{k}}$.
The Hamiltonian matrix $\mathcal{H}^{(q)}_{\bm{k}}$ is a generalization 
of (\ref{eq:ham}) and (\ref{eq:ham_bp}).
We define $q+2$ three-component vectors
$\bm{u}_{n,\bm{k}} = (u_{3n-2,\bm{k}},u_{3n-1,\bm{k}},u_{3n,\bm{k}} )^{\rm T}$,
with $n= 1,\cdots q +2$,
by which $\bm{U}_{\bm{k}}$ is written as $\bm{U}_{\bm{k}} = (\bm{u}_{1,\bm{k}}, \cdots, \bm{u}_{q+2,\bm{k}})^{\rm T}$.

Now, to obtain the flat bands, we assume the following:
the vectors $\bm{u}_{1,\bm{k}}$ and $\bm{u}_{q+2,\bm{k}}$, respectively corresponding to the wave functions on an upward triangle and a downward triangle,
satisfy the conditions,
\begin{equation}
[1,1,1] \cdot \bm{u}_{1,\bm{k}} =0, \label{eq:flatband_assumption}
\end{equation}
\begin{equation}
[e^{i \bm{k} \cdot \bm{a}_1},e^{i \bm{k} \cdot \bm{a}_2},1]\cdot \bm{u}_{q+2,\bm{k}} =0.
\label{eq:flatband_assumption2}
\end{equation}
These conditions are inferred from the wave function of a flat band of the kagome lattice~\cite{Hatsugai2011}, 
and ubiquitous among the class of models composed of 
interconnected triangles. 
From Eqs. (\ref{eq:flatband_assumption}) and (\ref{eq:flatband_assumption2}), one obtains
\begin{eqnarray}
h^{\bigtriangleup}  \bm{u}_{1,\bm{k}} = -  \bm{u}_{1,\bm{k}},  \label{eq:rel1}
\end{eqnarray}
and 
\begin{eqnarray}
h^{\bigtriangledown}_{\bm{k}} \bm{u}_{q+2,\bm{k}} = -  \bm{u}_{q+2,\bm{k}}. \label{eq:rel2}
\end{eqnarray}
Then, substituting (\ref{eq:rel1}) and (\ref{eq:rel2}) into (\ref{eq:sc}), 
we find that the Schr\"{o}dinger equation is reduced to 
the eigenvalue equation for the matrix whose dimension is $q+2$ as, 
\begin{eqnarray}
H^{\mathrm{C}}_{q}
\left(
\begin{array}{c}
u_{p,\bm{k}}\\
\vdots \\
u_{3(q+1) + p,\bm{k}}\\
\end{array}
\right)
= \varepsilon
\left(
\begin{array}{c}
u_{p,\bm{k}}\\
\vdots \\
u_{3 (q+1) + p,\bm{k}}\\
\end{array}
\right),
\end{eqnarray}
for $p = 1,2,3$,
where
\begin{eqnarray}
H^{\mathrm{C}}_{q}= \left[
\begin{array}{cccccc}
-t  & t_1  & 0 & \cdots & \cdots& 0\\
t_1 & 0 & t_2 & \cdots &\cdots & 0 \\
0 & t_2 &0& \cdots & \cdots& 0 \\
&  \vdots  & &\ddots& & \vdots  \\
0 & \cdots &\cdots &\cdots & 0 & t_{q+1} \\
0 & \cdots &\cdots & \cdots& t_{q+1} & -t \\
\end{array}
\right]. \label{eq:HamM}
\end{eqnarray}
Since $H^{\mathrm{C}}_{q}$ is independent of $\bm{k}$, we obtain $q + 2$ 
$\bm{k}$-independent eigenvalues, which are nothing but $q + 2$ flat bands.
It is worth noting that $H^{\mathrm{C}}_{q}$ can be regarded a 
tight-binding Hamiltonian of the $q+2$-site chain 
whose end points have the on-site energy $-t$ (Fig.~\ref{Fig:chain}).
It means that the flat-band energies can be obtained by calculating the eigenvalues
of the $q+2$-site chain; we emphasized that this is true
even when the object connecting the triangles is not a chain but a generic form of clusters. 

The wave function $\bm{U}_{\bm{k}}$ is obtained as follows:
Suppose that the eigenvector of $H^{\mathrm{M}}_{q}$ is $\bm{\psi}  = (\psi_1, \psi_2, \cdots, \psi_{q+2})^{\rm T}$.
Then, each component of $\bm{U}_{\bm{k}}$, $u_{i,\bm{k}}$,
is given by using three coefficients, $\lambda_1$, $\lambda_2$, and $\lambda_3$,
as 
\begin{eqnarray}
u_{3(n-1) + p ,\bm{k}}   = \lambda_p \psi_n,
\end{eqnarray}
with $n = 1, \cdots, q+2$ and $p= 1,2,3$.
Then, in order fort the conditions (\ref{eq:flatband_assumption}) and (\ref{eq:flatband_assumption2}) to be satisfied, 
$\lambda_p$ is determined as (up to the normalization constant),
\begin{eqnarray}
\left(
\begin{array}{c} \lambda_1 \\ \lambda_2 \\ \lambda_3 \\
\end{array} \right)
\propto \left(
\begin{array}{c} 1 \\ 1 \\1 \\
\end{array} \right)
\times 
\left(
\begin{array}{c} e^{i\bm{k}\cdot\bm{a}_1} \\ e^{i\bm{k}\cdot\bm{a}_2} \\1 \\
\end{array} \right)
= \left(
\begin{array}{c} 
1- e^{i\bm{k}\cdot\bm{a}_2} \\ 
e^{i\bm{k}\cdot\bm{a}_1} -1 \\ 
e^{i\bm{k}\cdot\bm{a}_2} - e^{i\bm{k}\cdot\bm{a}_1}\\
\end{array} 
\right).
\end{eqnarray}
\begin{figure}[!htb]
\begin{center}
\includegraphics[width= 0.5\linewidth]{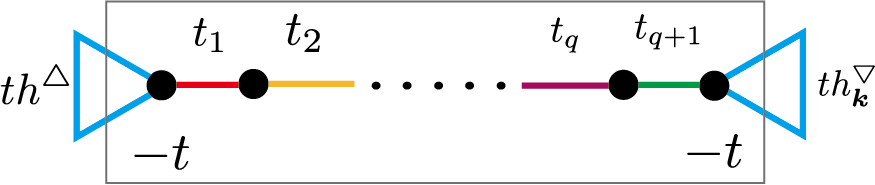}
\caption{Schematic figure for the $q + 2$-site chain (in a gray box)
described by the Hamiltonian of (\ref{eq:HamM}).
Blue triangles at the ends correspond to $th^{\bigtriangleup}$ and $th^{\bigtriangledown}_{\bm{k}}$,
operation of which gives rise to the on-site energy $-t$ at the end sites [see Eqs. (\ref{eq:rel1}) and (\ref{eq:rel2})].}
\label{Fig:chain}
\end{center}
\end{figure}

For a specific example of $q=1$ and $t_1 = t_2 = t^\prime$, 
which corresponds to Eq. (\ref{eq:ham}), 
the flat-band energies 
are given by the eigenvalues of the matrix
\begin{eqnarray}
H^{\mathrm{C}}_{1} = 
\left[
\begin{array}{ccc}
-t &t^\prime &0 \\
t^\prime &0 & t^\prime\\
0 & t^\prime &-t \\
\end{array}
\right],
\end{eqnarray}
that are $\varepsilon = -t, \frac{1}{2} \left( -t \pm \sqrt{t^2 + 8t^{\prime 2}}\right)$, as shown in Sec.~\ref{sec:model}.

\subsection{Explicit forms of the projection operators}
Here, we show the explicit forms of $\phi^{(j)}_{p,\bm{k}}$ of Eq. (\ref{eq:projection}).

Firstly, as for the flat band at $-t$, we have 
\begin{eqnarray}
\bar{\mathcal{H}}^{(1)}_{\bm{k}} = \mathcal{H}^{\rm P}_{\bm{k}} + t E_9 =
\left[
\begin{array}{c}
\phi^{(1)}_{1,\bm{k}}  \\
\phi^{(1)}_{2,\bm{k}}  \\
\phi^{(1)}_{3,\bm{k}}  \\
\phi^{(1)}_{4,\bm{k}}  \\
\phi^{(1)}_{5,\bm{k}}  \\
\phi^{(1)}_{6,\bm{k}}  \\
\phi^{(1)}_{7,\bm{k}}  \\
\phi^{(1)}_{8,\bm{k}}  \\
\end{array}
\right]
h^{(1)}_{\bm{k}}
\left[
\begin{array}{cccccccc}
\phi^{(1) \dagger}_{1,\bm{k}},
\phi^{(1)\dagger}_{2,\bm{k}} ,
\phi^{(1)\dagger}_{3,\bm{k}}, 
\phi^{(1)\dagger}_{4,\bm{k}},  
\phi^{(1)\dagger}_{5,\bm{k}},  
\phi^{(1)\dagger}_{6,\bm{k}},  
\phi^{(1)\dagger}_{7,\bm{k}}, 
\phi^{(1)\dagger}_{8,\bm{k}} 
\end{array}
\right], \label{eq:ham_proj_1}
\end{eqnarray}
where 
$\phi^{(1)}_{1,\bm{k}}  = (1,1,1,0,0,0,0,0,0)^{\mathrm{T}}$,
$\phi^{(1)}_{2,\bm{k}}  = (0,0,0,0,0,0,e^{-i \bm{k} \cdot \bm{a}_1},e^{-i \bm{k} \cdot \bm{a}_2},1 )^{\mathrm{T}}$
$\phi^{(1)}_{3,\bm{k}} = (1,0,0,0,0,0,1,0,0)^{\mathrm{T}}$, 
$\phi^{(1)}_{4,\bm{k}} = (0,0,0,1,0,0,0,0,0)^{\mathrm{T}}$, 
$\phi^{(1)}_{5,\bm{k}}  = (0,1,0,0,0,0,0,1,0)^{\mathrm{T}}$, 
$\phi^{(1)}_{6,\bm{k}}  = (0,0,0,0,1,0,0,0,0)^{\mathrm{T}}$,
$\phi^{(1)}_{7,\bm{k}} = (0,0,1,0,0,0,0,0,1)^{\mathrm{T}}$, 
$\phi^{(1)}_{8,\bm{k}} = (0,0,0,0,0,1,0,0,0)^{\mathrm{T}}$, 
and 
$$
h^{(1)}_{\bm{k}} =
\left[
\begin{array}{cccccccc}
t & 0 & 0 & 0 & 0 & 0 & 0 & 0\\
0 & t & 0 & 0 & 0 & 0 & 0 & 0\\
0 & 0 & 0 & t^\prime & 0 & 0 & 0 & 0 \\ 
0 & 0 & t^\prime & t & 0 & 0 & 0 & 0 \\ 
0 & 0 & 0 & 0 & 0 & t^\prime & 0 & 0 \\
0 & 0 & 0 & 0 & t^\prime & t & 0 & 0 \\
0 & 0 & 0 & 0 & 0 & 0 & 0 & t^\prime \\ 
0 & 0 & 0 & 0 & 0 & 0 & t^\prime& t \\ 
\end{array}
\right].
$$
From Eq. (\ref{eq:ham_proj_1}), we see the eight out of nine degrees of freedom are described by $\phi^{(1)}_{p,\bm{k}}$, 
therefore the remaining 1 mode, which is the kernel of $\Phi^{(1)}_{\bm{k}}$, has to be a zero-energy mode for $\bar{\mathcal{H}}_{\bm{k}}$. 
For the original Hamiltonian $\mathcal{H}_{\bm{k}}$, it corresponds to the flat band with the energy $-t$. 

Similarly, for the flat band at $-\frac{1}{2} \left( t + \sqrt{t^2 + 8t^{\prime 2}} \right)$,
we can rewrite the Hamiltonian as 
\begin{eqnarray}
\bar{\mathcal{H}}^{(2)}_{\bm{k}} = \mathcal{H}_{\bm{k}} + \frac{1}{2} \left( t + \sqrt{t^2 + 8t^{\prime 2}} \right)  E_9 
= \left[
\begin{array}{c}
\phi^{(2)}_{1,\bm{k}} \\ 
\phi^{(2)}_{2,\bm{k}}  \\ 
\phi^{(2)}_{3,\bm{k}}  \\ 
\phi^{(2)}_{4,\bm{k}}  \\ 
\phi^{(2)}_{5,\bm{k}}  \\ 
\phi^{(2)}_{6,\bm{k}}  \\ 
\phi^{(2)}_{7,\bm{k}}  \\ 
\phi^{(2)}_{8,\bm{k}}  \\ 
\end{array}
\right]
h^{(2)}_{\bm{k}} 
\left[
\phi^{(2) \dagger}_{1,\bm{k}},
\phi^{(2) \dagger}_{2,\bm{k}},
\phi^{(2) \dagger}_{3,\bm{k}},
\phi^{(2) \dagger}_{4,\bm{k}},
\phi^{(2) \dagger}_{5,\bm{k}},
\phi^{(2) \dagger}_{6,\bm{k}},
\phi^{(2) \dagger}_{7,\bm{k}},
\phi^{(2) \dagger}_{8,\bm{k}}
\right]
\end{eqnarray}
where $\phi^{(2)}_{1,\bm{k}} = (1,1,1,0,0,0,0,0,0)^{\mathrm{T}}$, 
$\phi^{(2)}_{2,\bm{k}} = (0,0,0,0,0,0,e^{-i \bm{k} \cdot \bm{a}_1},e^{-i \bm{k} \cdot \bm{a}_2},1 )^{\mathrm{T}}$, 
$\phi^{(2)}_{3,\bm{k}} = (x_1,0,0,x_2,0,0,0,0,0)^{\mathrm{T}}$, 
$\phi^{(2)}_{4,\bm{k}} = (0,0,0,x_2,0,0,x_1,0,0)^{\mathrm{T}}$, 
$\phi^{(2)}_{5,\bm{k}} = (0, x_1,0,0,x_2,0,0,0,0)^{\mathrm{T}}$, 
$\phi^{(2)}_{6,\bm{k}} = (0,0,0,0,x_2,0,0,x_1,0)^{\mathrm{T}}$, 
$\phi^{(2)}_{7,\bm{k}} = (0,0,x_1,0,0,x_2,0,0,0)^{\mathrm{T}}$, 
$\phi^{(2)}_{8,\bm{k}} = (0,0,0,0,0,x_2,0,0,x_1)^{\mathrm{T}}$, 
and
$$
h^{(2)}_{\bm{k}} = \mathrm{diag} 
\left[ t, t, 1,1,1,1,1,1 \right],
$$
with $x_1 = \sqrt{\frac{\sqrt{t^2 + 8t^{\prime 2}}-t}{2}} $ and $x_2 =  \sqrt{\frac{\sqrt{t^2 + 8t^{\prime 2}} + t}{4}} $.

Finally, for the flat band at $-\frac{1}{2} \left( t - \sqrt{t^2 + 8t^{\prime 2}} \right)$, we have 
\begin{eqnarray}
\bar{\mathcal{H}}^{(3)}_{\bm{k}}  \mathcal{H}_{\bm{k}} + \frac{1}{2} \left( t - \sqrt{t^2 + 8t^{\prime 2}} \right)  E_9 
= \left[
\begin{array}{c}
\phi^{(3)}_{1,\bm{k}}  \\ 
\phi^{(3)}_{2,\bm{k}}  \\ 
\phi^{(3)}_{3,\bm{k}}  \\ 
\phi^{(3)}_{4,\bm{k}}  \\ 
\phi^{(3)}_{5,\bm{k}}  \\ 
\phi^{(3)}_{6,\bm{k}}  \\ 
\phi^{(3)}_{7,\bm{k}}  \\ 
\phi^{(3)}_{8,\bm{k}}  \\ 
\end{array}
\right]
h^{(3)}_{\bm{k}} 
\left[
\phi^{(3) \dagger}_{1,\bm{k}},
\phi^{(3) \dagger}_{2,\bm{k}},
\phi^{(3) \dagger}_{3,\bm{k}},
\phi^{(3) \dagger}_{4,\bm{k}},
\phi^{(3) \dagger}_{5,\bm{k}},
\phi^{(3) \dagger}_{6,\bm{k}},
\phi^{(3) \dagger}_{7,\bm{k}},
\phi^{(3) \dagger}_{8,\bm{k}}
\right]
\end{eqnarray}
where $\phi^{(3)}_{1,\bm{k}} = (1,1,1,0,0,0,0,0,0)^{\mathrm{T}}$, 
$\phi^{(3)}_{2,\bm{k}} = (0,0,0,0,0,0,e^{-i \bm{k} \cdot \bm{a}_1},e^{-i \bm{k} \cdot \bm{a}_2},1 )^{\mathrm{T}}$, 
$\phi^{(3)}_{3,\bm{k}} = (y_1,0,0,-y_2,0,0,0,0,0)^{\mathrm{T}}$, 
$\phi^{(3)}_{4,\bm{k}} = (0,0,0,-y_2,0,0,y_1,0,0)^{\mathrm{T}}$, 
$\phi^{(3)}_{5,\bm{k}} = (0, y_1,0,0, -y_2,0,0,0 ,0)^{\mathrm{T}}$, 
$\phi^{(3)}_{6,\bm{k}} = (0,0,0,0,-y_2,0,0,y_1,0)^{\mathrm{T}}$, 
$\phi^{(3)}_{7,\bm{k}} = (0,0,y_1,0,0,-y_2,0,0,0)^{\mathrm{T}}$, 
$\phi^{(3)}_{8,\bm{k}} = (0,0,0,0,0,-y_2,0,0,y_1)^{\mathrm{T}}$, 
and
$$
h^{(3)}_{\bm{k}} = \mathrm{diag} 
\left[ t, t, -1,-1,-1,-1,-1,-1 \right],
$$
with $y_1 = \sqrt{2} x_2 $ and $y_2 = \frac{x_1}{\sqrt{2}} $.
\end{widetext}

\end{document}